# A holistic solution to icing by acoustic waves: de-icing, active anti-icing, sensing with piezoelectric crystals, and synergy with thin film passive anti-icing solutions

*Jaime del Moral,[+] Laura Montes,[+] Víctor J. Rico, Carmen López-Santos, Stefan Jacob,\* Manuel Oliva, Jorge Gil-Rostra, Armaghan Fakhfouri, Shilpi Pandey, Miguel Gonzalez del Val, Julio Mora, Paloma García-Gallego, Pablo F. Ibáñez-Ibáñez, Miguel A. Rodríguez-Valverde, Andreas Winkler, Ana Borrás\* and Agustín R. González-Elipe\**

([+]) These coauthors contributed equivalently

J. del Moral, L. Montes, V. J. Rico, C. López-Santos, S. Jacob, M. Oliva, J. Gil-Rostra, A. Borrás, A. R. González-Elipe

Nanotechnology on Surfaces and Plasma Lab, Materials Science Institute of Seville (CSIC-US), c/ Américo Vespucio 49, 41092, Seville, Spain.

C. López-Santos

Departamento de Física Aplicada I, Escuela Politécnica Superior, Universidad de Sevilla, C/ Virgen de Africa 7, 41011, Seville, Spain.

S. Jacob, A. Fakhfouri, S. Pandey, A. Winkler

Leibniz IFW Dresden, SAWLab Saxony, Helmholtzstr. 20, 01069 Dresden, Germany.

M. Oliva

Departamento de Física Atómica, Molecular y Nuclear, Avd. Reina Mercedes s/n, 41012, Seville, Spain.

M. Gonzalez del Val, J. Mora, P. García-Gallego

National Institute for Aerospace Technology (INTA), Ctra. Ajalvir km. 4, 28850, Torrejón de Ardoz, Spain.

P. F. Ibáñez-Ibáñez, M. A. Rodríguez-Valverde

Departamento de Física Aplicada, Universidad de Granada, Avd. de Fuente Nueva s/n, 18002, Granada, Spain

E-mail: s.jacob@ifw-dresden.de; arge@icmse.csic.es; anaisabel.borras@icmse.csic.es

Keywords: acoustic waves, de-icing, ice monitoring, freezing delay, ice-adhesion, ZnO, PFOTES

Icing has become a hot topic both in academia and in the industry given its implications in transport, wind turbines, photovoltaics, and telecommunications. Recently proposed de-icing solutions involving the propagation of acoustic waves (AWs) at suitable substrates may open the path for a sustainable alternative to standard de-icing or anti-icing protocols. Herein we unravel the fundamental interactions that contribute to the de-icing and/or hinder the icing on AW-activated substrates. The response toward icing of a reliable system consisting of a piezoelectric plate activated by extended electrodes has been characterized at a laboratory scale and in an icing wind tunnel under realistic conditions. Experiments have shown that the surface modification with anti-icing functionalities provides a synergistic response when activated with AWs. A thoroughtful analysis of the resonance frequency dependence on experimental variables such as temperature, ice formation, or wind velocity demonstrates the application in real-time monitoring of icing processes.

**1. Introduction**

Icing on surfaces is a common event hindering the development of both domestic and industrial activities and processes. Icing can be quite deleterious provoking accidents (e.g. in aviation or autonomous driving) or the malfunction of industrial and energy production facilities (e.g. wind turbines, marine structures, heat exchangers, and telecommunication antennas). Icing may also impact daily used instruments or vehicles (e.g., displays, outdoor sensors, cameras, and windscreens in cars among others). Therefore, a large variety of procedures have been developed and applied during the last years trying to mitigate the effect of icing in a large variety of scenarios. These can be classified into three main groups, namely, *anti-icing* (to avoid the formation of ice by applying surface engineering methods), *de-icing* (to remove ice once it is accreted on a surface), and combined *anti-icing* and *de-icing*. *De-icing* relies on distinct physical principles, including the application of low-temperature freezing liquids,[1] the use of the Joule effect,[2,3] solar actuation and photothermal effects,[4–6] the de-icing activation with plasmas,[7] the implementation of plasmonic nanoparticles,[8] the application of bulk or guided ultrasonic sound waves in the kHz range generated by external piezoelectric devices,[9] or, as reported very recently, the use of surface acoustic waves (SAWs) in suitable substrates coated with a piezoelectric thin film.[10]

In this article, we go a step forward and disclose basic ice-substrate interactions upon activation with bulk acoustic waves (AWs) generated in excited piezoelectric substrates. The investigation follows a holistic approach and studies *de-icing* processes of ice accumulated on surfaces, the promotion of an anti-icing function preventing the formation of ice (i.e. *active anti-icing*), the reduction of ice adhesion, and, interestingly, the possibilities to implement the proposed device as a smart sensor-actuator system (see **Scheme 1**). In general, AWs can be generated by the electrical activation of piezoelectric materials through the so-called inverse piezoelectric effect.[11] This can give e.g. rise to bulk or plate AWs, affecting in this case the whole structure and thickness of the material, or be confined to the surface, as for the so-called surface acoustic waves (SAWs).[12–14]

Although the incorporation of external piezoelectric devices onto a variety of materials has been claimed since some time ago as a suitable technological procedure to activate *de-icing* processes or to prevent the formation of ice on surfaces,[9,15] only recently it has been proposed the use of SAWs for de-icing[10] and/or for ice detection.[16–18] However, further analysis is required regarding aspects such as the *de-icing* capacity outside the area defined by the electrodes[10] and, particularly, the demonstration that AW *de-icing* can be used not only in the laboratory for small ice agglomerates but in real or close to real environmental conditions and large amounts of ice. In addition, in this work, we propose to advance into the combination of active and passive functionalities in the same device and exploit possible synergies arising from the incorporation of *anti-icing* thin films onto the piezoelectric substrates. More specifically, we address these issues to gain empirical knowledge about the effects of AWs on accreted ice, its formation on activated surfaces, and the operational possibilities of AW devices under realistic environmental conditions in an icing wind tunnel (IWT). Both *de-icing* of already formed ice and *active anti-icing* experiments (i.e. avoiding the formation of ice on continuously activated substrates) have been performed with LiNbO$_3$ piezoelectric plates. Specifically, shear modes of standing Lamb waves[19] were generated by applying an activation signal in the range of a few MHz and employing two planar electrodes rather than interdigitated electrodes (IDTs) as for SAWs.[10,12–14] Such a configuration is simpler for manufacturing, compatible with piezoelectric polymers and compatible with the realization of different experiments such as ice adhesion, ice accretion, or ice melting with the same device. A variable that has also been explored in this work is the effect of modifying the surface state of LiNbO$_3$ with anti-icing coatings and multilayers. We designate as *passive anti-icing* the function of some surface terminations preventing the formation of ice or repelling its accretion,[20–22] for example, by hydrophobic and superhydrophobic surface functionalization (although superhydrophobicity is not synonymous of an effective *anti-icing* function).[23,24] Herein, the modification of the surface state of LiNbO$_3$ has been carried out with a double perspective: on the one hand to make its surface hydrophobic (LiNbO$_3$ surface is hydrophilic) and, on the other, to prove whether the AW activation of the substrate remains effective to induce *de-icing* (i.e. to prove whether AW attenuation effects are negligible when a *passive anti-icing* layer is deposited onto the piezoelectric substrate material). For this purpose, we have deposited onto the LiNbO$_3$ a ZnO thin film, some hundreds of nanometers thick, which is further modified following a two-step approach: firstly by plasma deposition of a CF$_x$ thin film (i.e. Teflon like layer)[25,26] known for its hydrophobic and *passive anti-icing* properties and, secondly, by molecular grafting of a perfluorooctyltriethoxysilane (PFOTES) molecule, a surface termination that induces hydrophobicity and anti-icing surface responses.[27,28] ZnO is a typical piezoelectric material that, in the form of relatively thick films (generally over 5 µm), is widely used to generate SAWs with IDTs for a large variety of applications.[23,29,30] Finally, some comparative experiments have also been carried out with *black* LiNbO$_3$, a partially reduced variety piezoelectric material showing piezoelectric but not pyroelectric activity.[31]

Thus, as presented in Scheme 1, this work aims to demonstrate the suitability of AWs generated in piezoelectric plates to induce *ice-phobicity* and provide *ice-detection*. We trust that proving the effectiveness of the AWs de-icing concept in close to real scenarios should open the way for future industrialization and applications of evolved versions of systems based on this principle. In addition, it is expected that demonstrating the reliability of simple and non-expensive model AWs devices based on commercially available piezoelectric plates, extended electrodes, and of these systems after their modification with anti-icing layers, will provide the ever-growing anti-icing scientific community with an approachable methodology to advance in the development of a new generation of energy-efficient and smart de-icing systems.

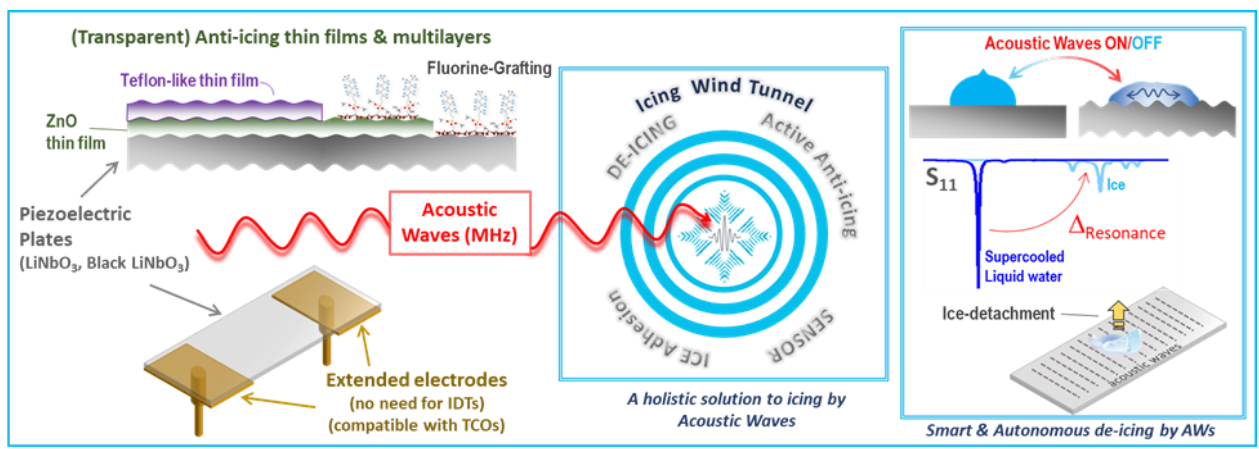

**Schematic 1.** Schematic representation of the scope of this article, from the application of piezoelectric plates and anti-icing thin films and multilayers to the holistic approach to the ice-phobic solutions.

## 2. Experimental Methods

### 2.1. Piezoelectric plate actuator and experiment design

The piezoelectric substrates utilized in the experiments consisted of double-side polished, single-crystalline $LiNbO_3$ plates (128º Y-cut, 0.5 mm thickness, 1x2 $cm^2$ or 1.5x3 $cm^2$ size, Roditi International Corporation Ld.). "Black" $LiNbO_3$ plates (Hangzhou FreqControl Electronic Technology) of the same size and crystal orientation cut were also used for dedicated experiments.

Rf electrical signals were applied onto two metal contacts 0.5 or 1 cm in width and 1 or 1.5 cm in length consisting of a Ti/Al bilayer of 5 / 295 nm thickness prepared by the lift-off technique and electron beam evaporation. Alternatively, contacts were also made by Ag layer electrodes (300 nm) prepared by DC-sputtering or simply by soft pressing a metal (Ta or Cu) foil (0.35 mm thickness, dimensions 0.5/1 cm) of the same dimensions located at the same position that the thin film contacts. No significant differences in AW actuation were found when comparing results obtained with these different types of electrodes.

To carry out the different *active anti-icing/de-icing* experiments, the plates working as chips were placed in a specifically designed and custom-built sample holder, which is presented in detail in the Supporting Information section, see Schematic 1 and **Figure S1**. A previous version of such a holder was developed by the authors to study the growth of $TiO_2$ thin films prepared by magnetron sputtering under acoustic wave activation.[32] In that work, we proved that Lamb type standing AWs are generated with this electrode-plate configuration. A characteristic feature of the holder system is that the piezoelectric plate is held just by two soft metal spring bolts (1 mm diameter with conical tip pressed onto the metal electrodes), and two holdings points or a thin bar at the lateral edges of the plate, respectively, to ensure that attenuation of the plate oscillation by the polymeric frame is minimized (see Figure S1). The electrodes are placed at the back of the plate, and thereby not exposed to the external icing environment. A reference sample (i.e. not subjected to AW activation) can be placed very close to the active piezoelectric plates for a proper evaluation of icing results, as discussed in the next section.

AW activation was carried out as indicated in the Supporting Information section **Figure S2**. The electrical system utilized for AW excitation of the piezoelectric substrates consisted of an rf signal generator (Agilent 8648A Synthesized RF generator, Keysight technologies), a high power amplifier ZHL-5W (Mini-Circuits), a vector network analyzer (VNA) (SDR-Kits, VNA 3SE), an oscilloscope (Siglent SDS 1204X-E) and a switching device enabling to connect the piezoelectric substrate either to the VNA or the rf signal source. The signal could be varied in voltage amplitude (from 10 to 50 V) and in frequency (around 3.4-3.5 MHz, a value around which it appears the resonance frequency mode of the plates used for the experiments). Under such experimental conditions, a standing Lamb wave excitation mode of AWs can be generated on the piezoelectric plate as already mentioned[32] and simulated in the present work (see below). The two electrodes on the piezoelectric plate were respectively connected to the rf signal source (i.e. function generator plus amplifier) and ground, in this latter case, intercalating a resistor to determine the $I(t)$ curves. $V(t)$ and $I(t)$ curves were continuously recorded with the oscilloscope connected as indicated in Figure S2.

A typical experimental protocol to control the generator signal consisted of the following steps:

1) Acquisition of the return loss spectrum (i.e. $S_{11}$ parameter) of a chip as a function of driving frequency. A typical example taken for a bare $LiNbO_3$ substrate chip is shown in **Figure S3**. It can be realized in this figure that each plate can be characterized by a single or various resonance peaks in the 3.4-3.6 MHz range. For the experiments, the function generator is set at the frequency of the highest minimum of the return loss spectrum (in the example of Figure S3 at 3.5609 MHz).

2) Application of the rf excitation signal at the set frequency and follow the icing/de-icing event of interest.

3) Dynamic re-adjustment of the rf excitation frequency at the $S_{11}$ minimum after the change of acoustic load, e.g. due to freezing or droplet aerosolization. Determination of changes in the $S_{11}$ minima was done by intermittent switching between VNA and the signal source.

**2.2. Simulation of generated AWs**

A numerical study was carried out to further understand the complex vibration patterns at the plate resonance. A $LiNbO_3$ plate with electrodes (without holder and circuit board) was modeled using a predefined set of three-dimensional coupled electrodynamic and mechanic equations in the commercial finite element (FE) solver COMSOL Multiphysics version 6.0 (similar to the numerical setup described in ref. [32]). The geometry of the plate had the same dimensions as the devices used in the experiments. The bulk material was described with an anisotropic, linear-elastic, and piezoelectric material model, using the stiffness, piezoelectric, and permittivity tensors for transparent $LiNbO_3$ depicted in ref. [33]. Preliminary computations at the design frequency did not show a noticeable effect of the mass and stiffness of the aluminum electrodes on the observed vibrations (not shown). Therefore, the electrodes were simplified as two-dimensional, massless, and soft patches at the surface of the bulk material. The plate was discretized with a rectilinear grid with first-order shaped elements, consisting of [40650] nodes, resulting in [928144] degrees of freedom. Computations at the design frequency with different mesh densities showed the independence of the results from the resolution of discretization (not shown). The mechanical boundary conditions mimicked the mounting of the device during the experiment, as in Figure S1, prohibiting displacement normal to the substrate surface at the connector pins and the outer chip surface opposing the electrodes. The electrical boundary conditions prescribed time-harmonic electric potentials uniformly distributed on the electrode patches. The frequency of the time-harmonic signal was varied from 3.4 MHz to 3.7 MHz in 1 KHz steps, and the model was solved for each of the steps.

### 2.3. Surface modification with anti-icing grafting and multilayers

Icing tests were carried out either using the bare $LiNbO_3$ substrates or onto these substrates functionalized by the deposition of various thin layers modifying the wetting and *passive anti-icing* response of the examined substrates. The following surface terminations have been applied in this study:

*ZnO thin films.* Polycrystalline ZnO thin films were deposited by plasma enhanced chemical vapor deposition (PECVD) at room temperature in an ERC MW plasma reactor as detailed elsewhere.[34–36] Typical thicknesses of the ZnO thin films deposited onto the $LiNbO_3$ substrate varied between 500 to 1500 nm, a range for which no significant differences were found in the *de-icing* and *active anti-icing* experiments.

*$CF_x$, Teflon like thin films.* $CF_x$ thin films have been prepared at ambient temperature by PECVD in an rf parallel plate reactor. Full details of the preparation procedure of these thin films have been reported elsewhere.[25,26] $CF_x$ thin films with thicknesses in the range of 100 to 200 nm were deposited directly onto the $LiNbO_3$ plates or onto ZnO coated plates. Hereafter, this bilayer arrangement will be named $ZnO/CF_x$.

*PFOTES molecular grafting.* The bare or ZnO-coated $LiNbO_3$ plates were functionalized by the grafting of perfluorooctyltriethoxysilane (PFOTES) molecules following the procedure detailed in refs. [28,37]. The grafted ZnO layer will be labeled as ZnO(F) and the bare substrate as $LiNbO_3$(F).

The morphology of these thin films was characterized by Scanning Electron Microscopy (SEM) in an S4800 FESEM microscope by Hitachi, working at 2 kV and a typical working distance between 2 and 3 mm. The chemical state of the functionalization films has been determined by X-ray photoelectron spectroscopy (XPS) in a a SPECS PHOIBOS-100 spectrometer operated with Al Kα radiation as excitation source and 20 eV constant pass energy.

### 2.3. Icing Tests

Specially designed experiments have been carried out to get information about the effect of AW activation on the following icing issues and interactions: i) droplet icing and melting; ii) ice adhesion; iii) ice accretion, under the premises of *de-icing* and *active anti-icing* processes, as described in the following:

*i) Droplet icing and melting.* These experiments included the deposition of water droplets onto the piezoelectric substrate, their freezing upon cooling, and their subsequent melting by AW activation. The analysis has been carried out either in a cooling chamber or on a Peltier plate emplaced in the controlled temperatura chamber of an OCA20 goniometer (Dataphysics). The temperature was decreased from ambient temperature to values between -5 and -20 °C. Enlarged images of the water droplets, taken with a photo camera in cross-section view, have been analyzed to disclose the basic mechanism of melting while activating the substrate with AWs. In parallel, the $S_{11}$ parameter was recorded with the VNA as a function of temperature and when the melting events are taking place. It will be shown that these spectra provide information on the actual temperature of the piezoelectric plates, as well as on the freezing/melting process of the water droplets.

*ii) Ice adhesion tests.* Ice adhesion tests were carried out by the pull-off method described in detail in reference.[38] The experimental set up including a description of the pulling probe attached to the piezoelectric device ready for activation is reported in the Supporting Information section, **Figure S4**. The experiment proceeds by emplacing a hollow Teflon cylinder on the sample surface and filling it with 1 mL of Milli-Q water (cylinder's internal diameter is 9.86 ± 0.12 mm, its area 76.4 ± 1.9 mm$^2$, and the water level is about 13 mm). Then, the sample was placed in a freezing chamber and left to freeze at a temperature of -10 °C, waiting until full solidification is reached. Once the ice was formed, a pulling force was applied through a thread connected to a dynamometer (IMADA ZTA-200N/20N), which placed outside the cooling chamber was attached to a motorized linear stage (IMADA MH2-500N-FA). The thread was subjected to a progressively increasing force until the ice probe became released from the surface. The applied force was perpendicular to the sample (tensile mode of analysis) and was continuously read by the dynamometer. It can be directly converted into adhesion strength values by referring the value of the force producing the detachment of ice to the area of the probe, (i.e. Force(N)/Area(m$^2$), magnitude expressed in stress units of MPa). The setup was adapted to make it compatible with AW *de-icing* conditions. Different experiments were carried out under passive conditions (i.e. without applying rf signal to the chip), continuous rf actuation of the chip for a defined voltage, and the combination of both (i.e., activating the chip for a given time and then applying the force without rf activation).

*iii) Ice accretion, de-icing and active anti-icing in IWT*. Ice accretion experiments were carried out to test the *de-icing* and *active anti-icing* performance of piezoelectric substrates under AW activation. Tests were carried out in an IWT located at INTA (Madrid, Spain) and described elsewhere.[28,38,39] Experiments were done using liquid water contents (LWC) in the range 0.18 – 0.22 g/m$^3$, supercooled droplets with 20 µm of median volume diameter (MVD) in size distribution, a temperature from -5.5 to -8.5ºC, and three wind tunnel speeds: 25, 50, and 70 m/s. Ice accretion was monitored simultaneously on AW-activated LiNbO$_3$ substrates and the reference non-actuated surface placed beside. To control the arrival direction of the air and water flow onto the plates a screen frame with two slits of 67 mm$^2$ was placed before them at a distance of 19 mm. The geometry of this screen setup and related experimental details are presented in Figure S5. The evolution of the ice accretion on the surface of the reference and AW activated piezoelectric plates was followed by direct video imaging of their surfaces. The evolution of temperature at the surface of the AW activated chip plate during the experiments was monitored with a Testo 890 thermographic camera and, close to the chip, with a Fiber Bragg Grating Sensor (FBGS®). The optical fiber polyimide coating was previously removed to avoid decalibration, and replaced with a polyimide isolation capillary. The measurements were done using a LUNA Si155 Interrogator. The optical fiber was calibrated using SIKA TP3M165E2 highly precise temperature calibrator and a cubic calibration curve.[40]

## 3. Results and discussion

### 3.1. Simulation of AWs

We have recently shown that the excitation configuration proposed in Scheme 1 and Figures S1-S2, consisting of two planar electrodes subject to rf signaling around 3.5 MHz on LiNbO$_3$ plates, generates patterns of acoustic standing waves[32] with a submillimeter wavelength. Since the *de-icing* and *active anti-icing* processes studied in this work can be quite dependent on the AW characteristics the generated standing AWs have been simulated using the procedure described in the experimental and method section.

As in the experiments, we found a narrow-banded drop in the return loss computed at each simulated frequency, which indicated a strong coupling between the electric excitation and the AW mode at that frequency. The resonance frequency of this coupling was identified as 3.509 MHz (**Figure S6**), in good agreement with the experimental results. The deflection of the chip at resonance is presented in **Figure 1**a). Here snapshots during a period of a single vibration (T) are presented. An animation of a full cycle can be found in **Video S1**. Note, that the displacement of the surface is exaggerated (by a factor of 3000). The colors of the figure give a more realistic view of the vibration's magnitude, which was in the order of tens of nanometers (for 1W excitation power). The most important finding is the large in-plane displacement, suggesting a shear-dominant AW-mode. Figure 1b) presents the three displacement components along the centerline of the plate surface. We found that the shear component of the displacement was one order of magnitude larger than its normal component. Furthermore, the excited AW-mode resulted to be the first thickness-shear mode of the plate with an acoustic standing wave in the direction of the plate thickness. The frequency of this resonance depends on the material properties and the thickness of the substrate, but barely on the other dimensions of the device. This was confirmed in the experiments, where chips with different surface areas and electrode dimensions (but identical thickness) had the same resonant frequency. The findings suggest that the normal displacement of the chip surface (observed numerically, but also e.g. in vibrometer tests in ref. [32]) is a disturbance of the fundamental shear AW mode, rendered by the mechanical boundary conditions and the finite plate dimensions. It is noteworthy that the simulation only gives a qualitative assessment, as the true boundary conditions are unascertainable from the conducted experiments.

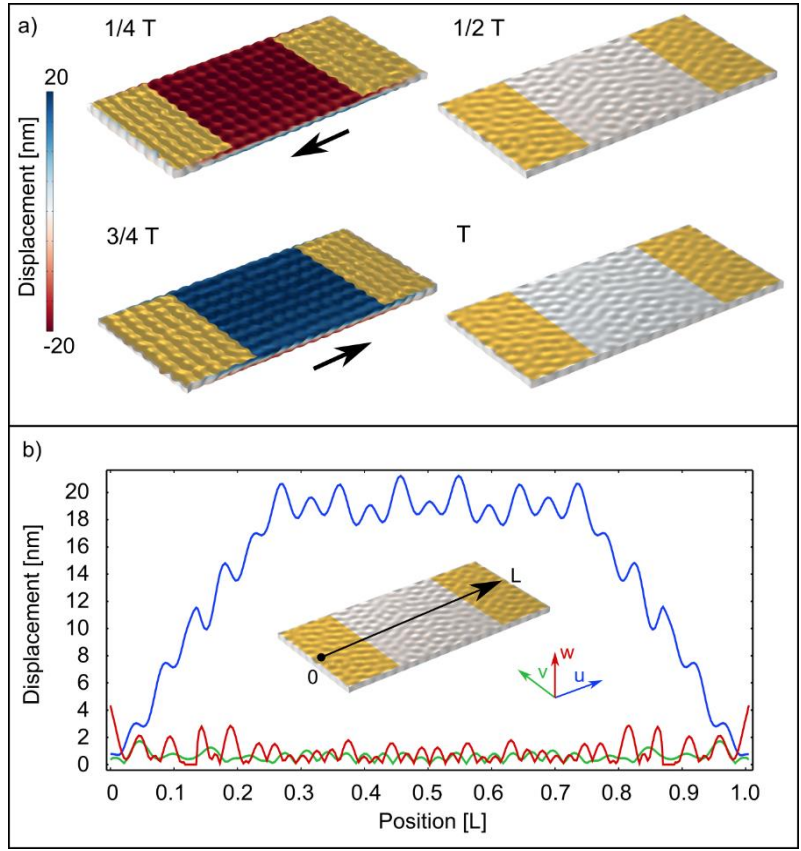

**Figure 1. Vibration of the LiNbO₃ plate with planar electrodes, computed with the FE method at the designed frequency.** a) Snapshots of the displacement fields (exaggerated with a scaling factor of 3000) at different fractions of a single period of the vibration (T). b) Absolute values of the displacement fields along the center-line, revealing a dominating share mode. An animation of the vibration is included in the supplementary material of the paper (Video S1)).

In the next sections we will describe the observed effects of these AWs on the specific icing experiments i)-iii), which, carried out as explained in the experimental section, should be primarily considered as resulting from the shear excitation mode of the induced AW. It must be remarked that other AW modes might deliver slightly different effects and efficiency, although we expect that basic mechanisms should be equivalent to those reported here.

### 3.2. Melting and de-icing of ice aggregates

The first step (i) of this investigation consisted of determining the influence of various experimental variables on the melting of water and ice droplets upon AW activation of a piezoelectric substrate. We examined the following variables: a) effect of temperature on the resonant conditions of the plates; b) idem by dripping water onto the surface and the formation of ice aggregates upon cooling; c) effect of the AW excitation of the piezoelectric substrate inducing *de-icing*. Herein, we will address these issues considering three types of piezoelectric substrates: bare substrates with an intrinsic hydrophilic wetting behavior, surface modified substrates with a hydrophobic/anti-icing behavior, and black non-pyroelectric LiNbO$_3$. Previous to reporting the results of the studied icing processes, we describe selected properties of surface-modified LiNbO$_3$ chip plates.

*3.2.1 Hydrophobically and anti-icing treated LiNbO$_3$ plates.*

To render the LiNbO$_3$ plate surface hydrophobic, we used ZnO thin films fabricated by PECVD. Although these ZnO thin films are relatively compact, their surface is known to present a certain roughness,[33,34,36] a feature that according to the Wenzel wetting model[41] will contribute to increasing the surface hydrophobicity and/or serve as suitable anchoring layer of water repellent molecules. The basic characteristics of the films used in the present study are shown in **Figure 2**. Figures 2a) and b) show normal and cross-section micrographs of a ZnO film illustrating its microstructure and surface morphology. ZnO thin films prepared by PECVD were crystalline and depicted a preferential crystallographic orientation of the (101) planes parallel to the surface.[27] It is noteworthy that fluorination with the PFOTES molecules (see below) did not induce appreciable changes in the microstructure or crystallinity of the films. In addition, the inset in Figure 2b) confirms the high transparency of the ZnO thin films, a requirement for the future implementation of AW activation procedures onto transparent devices (e.g. for display or photovoltaic applications).

Surface functionalization consisted of the fabrication of a ZnO/CF$_x$ bilayer, or the grafting of PFOTES molecules to get a ZnO(F) layer (see experimental section). The grafting process used to prepare the ZnO(F) layer entails the reaction of the methoxy groups attached to the silicon atom of the grafting PFOTES molecule with –OH groups at the surface of the oxide. In this regard, the higher surface roughness of the ZnO thin films with respect to the LiNbO$_3$ plates is favorable to enhance the grafting capacity of the outer surface of the devices and, therefore, increasing their hydrophobic and *anti-icing* capacity.[42] **Figure 2c)** shows a typical SEM cross-section view of a ZnO (800 nm)/CF$_x$ (300 nm) bilayer (see **Figure S7** for the top-view SEM image). This cross-section highlights the good definition and flat character of the interfaces existing between LiNbO$_3$ and ZnO, as well as ZnO and CF$_x$, where no large voids, imperfections, or defects can be devised. The planarity of interfaces is a requirement for a straightforward transmission of AWs through multilayer structures.[43,44] X-ray photoelectron spectroscopy analysis confirmed that the CF$_x$ films presented a PTFE-like composition where CF$_x$ chains with CF$_3$, CF$_2$, and CF functional groups constitute their basic structure[25,26] (see Supporting Information Figure S7). In previous works, we have demonstrated that these films are hydrophobic, conformal to rough surfaces, and may induce a Cassie-Baxter wetting regime,[45] as well as a robust *passive* anti-icing response.[42,46,47]

Both the ZnO/CF$_x$ and ZnO(F) device terminations showed a hydrophobic behavior. The wetting contact angle at room temperature determined for a 5 µl water droplet placed onto the LiNbO$_3$+ZnO/CF$_X$ and LiNbO$_3$+ZnO(F) surfaces was 110º and 133º, respectively, compared to 55º for the bare LiNbO$_3$ surfaces. The sliding angle (i.e. the tilting angle of the plate required for the drop to roll-off) was always higher than 90º in the two cases. These surface characteristics are expected to modify the interface interaction between substrate and water droplets during the freezing and melting processes.

Regarding the interaction of AWs and ice, a crucial issue is to verify that the electromechanical properties are not significantly affected by the coating functionalization, particularly because the non-ideal behaviour of the polycrystalline films as compared with the crystalline character of the substrate plate might induce a certain AW attenuation (see next sections showing that, effectively, resonance frequency was quite sensitive to the environment, amount of water or ice, temperature, and similar parameters). To discard significant damping or alteration of the AWs upon device functionalization, return loss spectra were recorded for the bare LiNbO$_3$ plate and the same plate with a ZnO film deposited onto its surface. The corresponding spectra reported in **Figure 2d)** are practically identical (except for a small frequency shift that could be attributed to small differences in the fixing of the plates to the sample holder), thus sustaining that, from the point of view of AW activation, the plates remain practically unmodified after the deposition of the ZnO thin film. The same behavior was found for LiNbO$_3$-ZnO/CF$_x$ and ZnO(F) device chips.

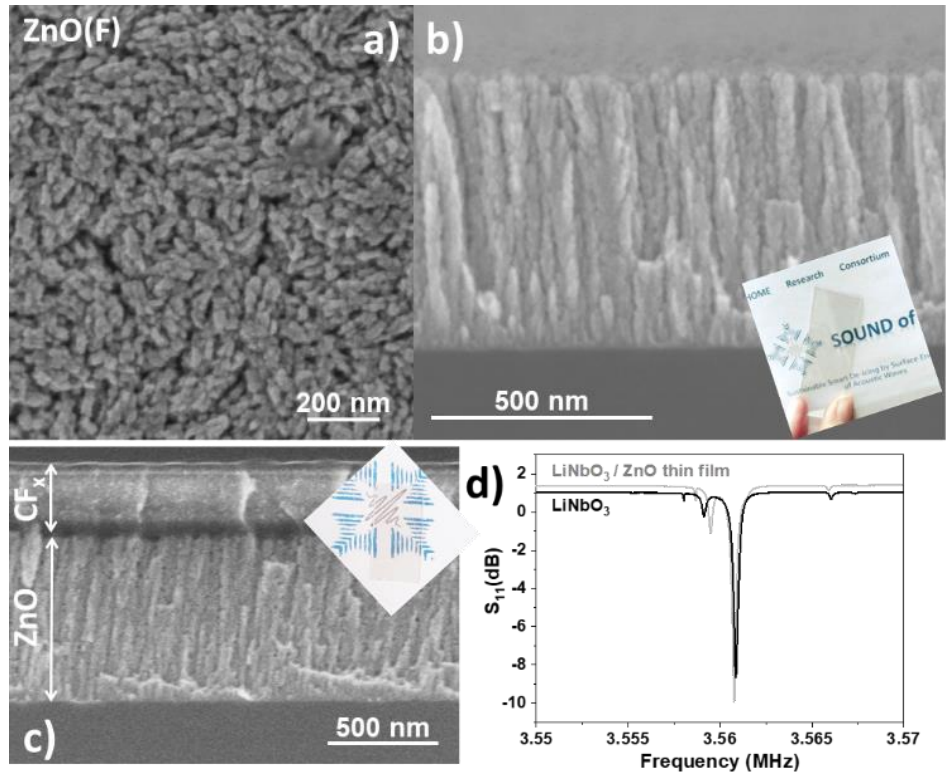

**Figure 2. Thin film layers and multilayers fabricated by plasma assisted methods as topcoats of bulk piezoelectric plates.** SEM micrographs of the perfluorinated ZnO (a-b) and a $CF_x$ polymeric thin film on ZnO (c). Insets in b) and c) demonstrate the high transparency of the layers and multilayers deposited on the piezoelectric plates. d) Comparison of the return loss spectra for the bare and modified $LiNbO_3$ plates as indicated in the plot.

*3.2.2 LiNbO$_3$ plate resonance and freezing of water droplets.*

A series of essays were firstly carried out to monitor the evolution of the plate resonant behavior through the typical steps of a water droplet freezing experiment,[28,48–51] namely: 1) dripping water on the surface at Temperature > 0 ºC, 2) progressive rapid cooling to the desired minus zero temperature and 3) verification of freezing of the water droplet by optical analysis.

It is noteworthy by this experimental sequence that step 3) may occur at temperatures well below zero (i.e. water remains in a super-cooled state) and take a considerable time due to various freezing delay factors.[52] **Figure 3** shows the evolution of the return loss spectra recorded when varying the temperature for a series of $LiNbO_3$ plates without (i.e. dry) and with a 5 μl water droplet deposited onto their surface (Figure 3 a) and b), respectively). Figure 3a) shows the series of spectra recorded for the dry plate decreasing the temperature from 18 ºC down to -20 ºC. Figure 3c) shows that a linear shift in the position of the return loss minimum occurs with temperature. This dependence was completely reversible upon increasing the temperature and is characterized by a slope of -85 ppm/ºC, in good agreement with reported values (-86.44 ppm/°C) of the temperature coefficient constant of $LiNbO_3$.[53–55] Figure 3b) also gathers an equivalent series of $S_{11}$ spectra recorded for the same plate but with a 5 µl droplet deposited onto the surface. Despite the wider shape of the signal, attributable to the presence of the water droplet on the surface of the chip, the $S_{11}$ return loss signal experienced a similar frequency shift with temperature, provided that the water droplet remained liquid or in a supercooled state. As evidenced in Figure 3c), a drastic increase shift in resonance takes place upon water freezing. From a practical perspective, the similar linear variations irrespective of the presence of a water droplet on the surface suggest the possibility of extending the use of $LiNbO_3$, already utilized for high-temperature monitoring,[55,56] as a temperature sensor under humid and low-temperature conditions. The capacity to detect icing processes and to monitor the water-ice conversion is an additional function implicit in Figure 3b) showing that freezing of deposited water not only induces a sharp increase in resonance frequency of the $S_{11}$ signal but also changes its shape and intensity. From a fundamental point of view, these changes at around -20 ºC, when the supercooled water droplet transforms into an ice aggregate, point to a drastic modification of the electromechanical properties of the device. Such a result agrees with recent evidence by Anisimkin et al.[18] showing that an AW $LiNbO_3$ plate ice sensor operated with IDTs presented a much higher attenuation coefficient in the presence of surface ice than with water.

It is important to stress, that we found a similar evolution with temperature when comparing the results obtained with the bare $LiNbO_3$ substrate plate with those covered with an anti-icing coating, i.e., for samples $LiNbO_3$-ZnO(F), $LiNbO_3$-ZnO/$CF_x$ or $LiNbO_3$/$CF_x$ (Figure 3c). Similarities also exist with respect to the sharp increase in frequency accompanying the water-ice transformation (Figure 3c). This similar behavior indicates that the found variation with temperature responds solely to the effect of the temperature coefficient constant of $LiNbO_3$,[53,55] an effect agreeing with the predominant shear character of the AWs generated in the chip device utilized for the experiments. It is also noteworthy that the observed reproducible dependence of the $S_{11}$ resonance with temperature was also found for black $LiNbO_3$ (see Supporting Information **Figure S8**), a feature supporting that the variation of electromechanical properties with temperature are equivalent for these two types of $LiNbO_3$ plate substrates.

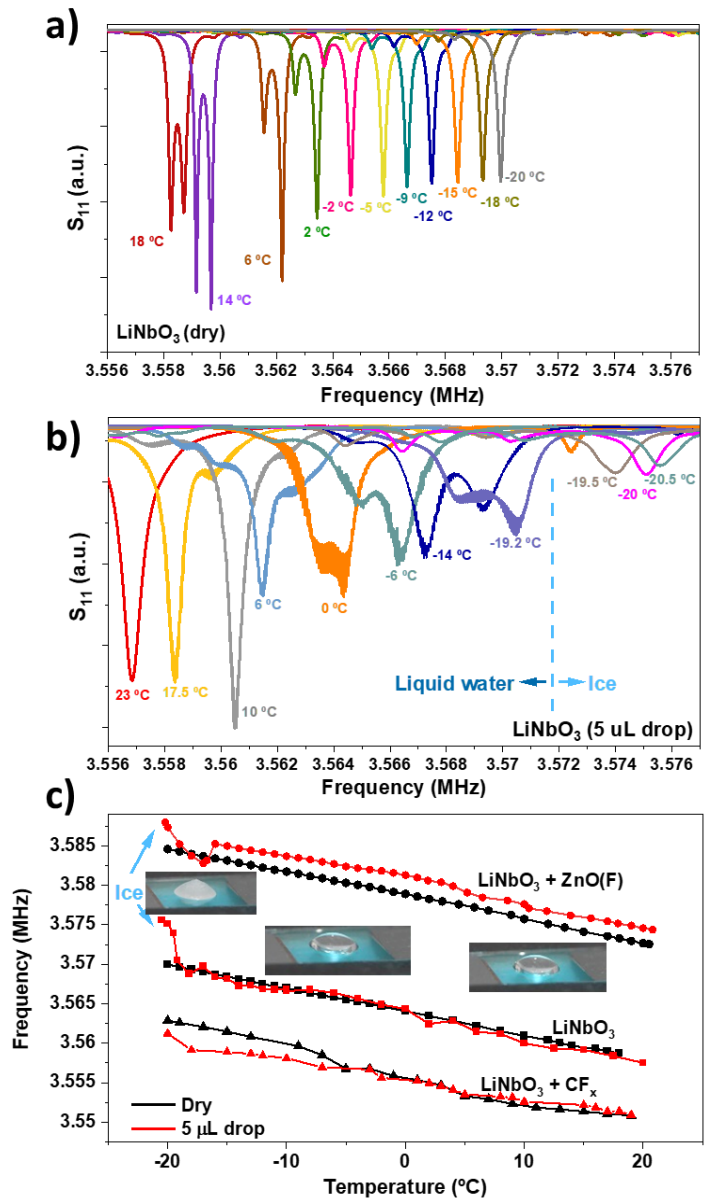

**Figure 3. Variation of return loss spectra of bare and modified LiNbO$_3$ plates with planar electrodes as a function of temperature and in the presence of water droplets and ice aggregates.** a) Return loss spectra (i.e. S$_{11}$ coefficient) of the LiNbO$_3$ plate as a function of frequency for decreasing temperatures from 18°C to -20° C. b) Idem for this plate with a 5µl water droplet on the surface. The drastic attenuation and shape change of the S$_{11}$ spectrum after freezing respond to the transformation of water into ice. c) Plot of the frequency of S$_{11}$ minimum against temperature for the bare and modified LiNbO$_3$ plate chips with and without water droplet (5 µl) on their surface. The arrows signal the points at which the super-cooled water droplet transforms into ice. The insets show photographs of a characteristic water droplet before and after its transformation into ice.

*3.2.3 Ice melting with AW activated LiNbO₃ plates.*

The previous analysis of the variation of LiNbO$_3$ return loss spectra with temperature and the formation of ice on its surface has revealed a high dependence of the resonance frequency with temperature and the formation of ice. It also indicates that an efficient ice removal via plate wave excitation on these substrates would require precise and continuous tuning of the resonance frequency during the *de-icing* process (see a detailed description of the tuning procedure in the Experimental Section).

Melting tests of ice aggregates were firstly carried out on a bare LiNbO$_3$ substrate. The time evolution of the ice aggregate profile upon AW activation, determined from the side image of the droplet recorded with a photo camera, is depicted in **Figure 4**a)-left. These profiles are compared with those observed during a de-icing test induced by slow heating up to ambient temperature (the complete melting sequence can be followed in **Video S2**). A first remarkable finding is the short time required to achieve complete melting, 41 s, by the AW activation with a 40 V rf excitation signal. It is also noteworthy that the aggregate shape during the melting process induced by AWs differs substantially from the profile found when performing the melting by natural heat (Figure 4a)-right). In particular, it appears that in intermediate (e.g. see snapshot at 25 s in Figure 4a-left) and final states (snapshot at 41 s) liquid water spreads onto the surface to a larger extension than for the warmed-up ice aggregate (e.g. the water droplet profile in Figure 4a-right after complete melting defines a wetting contact angle of ca. 10°, against ca. 30° for the droplet resulting from natural melting, the complete-time elapse is ca. 5 min 50 s (see Video S2). The complex mechanism underlying this effect, the so-called "acoustowetting", has been studied extensively in recent years, e.g. by Manor et al. [57–59] These results also suggest that during AW activation, water forms and accumulates at the outer surface of the ice-drop, including also the ice-substrate interface and, upon progressive melting, spreads onto the entire drop volumen and the substrate surface. The interface melting as an intermediate step in the AW *de-icing* mechanism agrees with a dominantly shear wave-driven melting mechanism and is further supported by the experiments in Figure 4b) and **Video S3.** The former shows a series of photographs taken during an AW-induced *de-icing* experiment with the substrate and ice aggregate placed vertically. The series of snapshots presented in this figure reveals that at intermediate states of the de-icing process, the remaining ice spreads over the surface and tends to move outside the image frame. This spreading indicates a limited adhesion to the substrate, in agreement with the formation of a water interlayer and the action of the gravitational force. The cartoons in Figure 4c) reproduce, in an ideal way, the observed evolution of the ice aggregate under AW activation for 41 sec. They represent a progressive melting of the ice aggregate at the interface, the spreading of water both onto the substrate and over the remaining core of ice (note that water contact angle on ice is very small if not zero and can easily spread over its surface).[60,61] Regarding the efficiency of the AW transmission, those conditions where water and ice coexist on the surface would likely release the attenuation in the AW transmission found after the formation of ice (i.e. Figure 3b), provided that the substrate contacts an interlayer of liquid water and not directly ice. This stems from the high effectiveness of AWs to energize water, as widely studied for the AW propagation in water droplets.[62,63] Interestingly, black $LiNbO_3$ behaves similar to the standard $LiNbO_3$ regarding the activation of the melting process of ice aggregates, as illustrated by the images provided in **Figure S9**. This sequence of images was taken along an experiment similar to that in Figure 4 and the results suggest that electromechanical activation processes rather than a pyroelectric effect are the main cause of the AW-induced melting of ice.

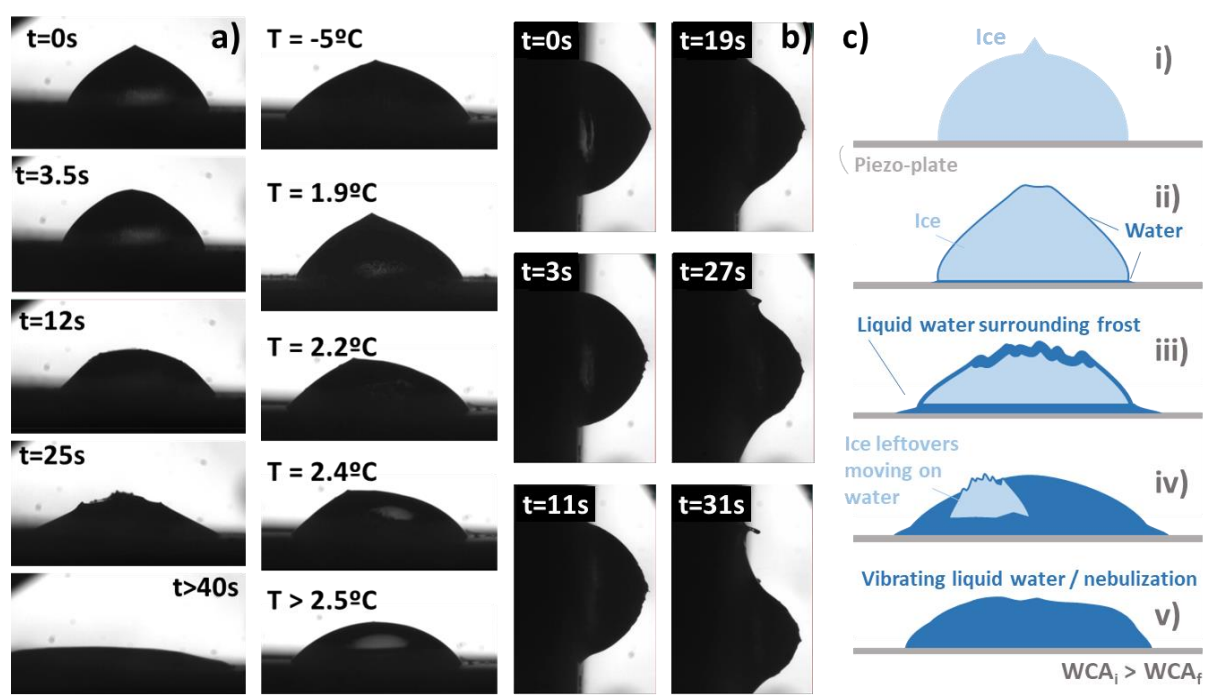

**Figure 4. Ice melting process mechanism through the analysis of profile snapshots of ice aggregates onto $LiNbO_3$ substrates.** a) Comparison of ice melting process from an ice aggregate excited with AWs (left) and by natural warming up (right). b) Ice melting process induced by AWs for the substrate placed vertically. c) Cartoon representing the ice/water evolution during the AW-induced ice melting process.

### 3.3. Ice adhesion tests

Ice adhesion strength on substrates is a fundamental interface property to achieve an efficient *anti-icing* response of solid surfaces.[38,46,64,65] Within a holistic approach, both surface chemistry and roughness should affect ice adhesion. Regarding the latter, it is usually admitted that the rougher the surface the higher the ice adhesion force between ice agglomerates and substrates.[66] The experiments gathered in this section aim at determining the effects of AW interaction on ice adhesion.

Essays have been carried out with the bare (and therefore sub-nm flat) $LiNbO_3$, perfluoro molecule-functionalized $LiNbO_3$ ($LiNbO_3(F)$) and the $LiNbO_3$ covered with anti-icing coatings (i.e. $LiNbO_3$ + ZnO(F) and $LiNbO_3$ + $ZnO/CF_x$). Such modifications convert the piezoelectric plate surface into hydrophobic but, as discussed in the previous sections, at the expense of incorporating a certain surface roughness, particularly in case of ZnO coating. Ice adhesion experiments on the different devices were carried out following the experimental protocol described in the experimental section, either without or with AW activation, and using a vertical pulling-off configuration. RF amplitude voltages of 20 and in the range 35/40 V were applied to the chip electrodes while pulling with an increasing force until the complete detachment of the ice probe. These AW excitation conditions did not induce the ice melting in the cantilever, although a limited interface melting or softening should be expected. It is also noteworthy that the application of a pulling force to the ice probe adhered to the surface may modify the resonance conditions of the piezoelectric device, e.g. by mechanical deformation of the substrate. This is evidenced in **Figure 5a)** comparing $S_{11}$ spectra taken while applying either zero or a net pulling force at the same temperature to ensure that chip resonance variation was only related to the pulling action and not to the evolution with temperature (cf. Figure 3). The shift in resonance frequency highlights the high sensitivity of the system to the environmental conditions and the need to finely tune the frequency of the applied AW excitation signal while performing the ice adhesion tests (note that a systematic analysis of return loss spectra as a function of the pulling force might be used to derive pressure sensing information with these devices.[38] This adjustment was systematically done before performing the actual adhesion tests under AW activation. The characteristic force/time curves recorded during ice adhesion tests experiments are shown in Figure 5b) for sample $LiNbO_3$+ZnO(F) while applying a voltage of 0 V (i.e. passive conditions) or 35 V. The two curves are characterized by a progressive increase in the pulling force with time as resulting from a progressive upward movement of the dynamometer that pulls off the ice probe. This continuous increase is suddenly stopped when the ice probe becomes detached from the surface. The maximum recorded value corresponds to the force required to detach the probe and can be used to estimate the adhesion strength under each working condition. The first result of the experiments was that the force required to detach the probe was significantly smaller when AWs were generated in the substrate during the detachment experiment (see values of adhesion strength in **Table 1**). In a recently published ice adhesion study using a shear actuated device under AW activation, authors observed a decrease in ice adhesion attributed to an increase of dipole spacing at the interface provoked by the nanoscale surface vibration of the piezoelectric.[64] In addition to this or similar interface effects, the experimental configuration utilized in this work evidences that the interface softening induced by the AW activation favors the complete detachment of the ice and its separation from the surface.

Moreover, Figure 5c) addresses the combination of active and passive working conditions. In this experiment, the rf excitation (at 35 V) was applied on the chip only for 5 seconds, immediately after, the pulling was started, obtaining an extremely low value of 2.84 N. Such a result indicates that the ice-adhesion is reduced already during the very first seconds of the actuation producing an effective softening of the ice interface with the piezoelectric plate. This working mode may also reduce the energy consumption in the eventual application of this methodology in a real scenario where the detachment of the ice could be induced under sequential pulses and even driven by self-deicing mechanisms, i.e. ice falling by its weight, splitting into debris, or detached by neighboring droplets.[67]

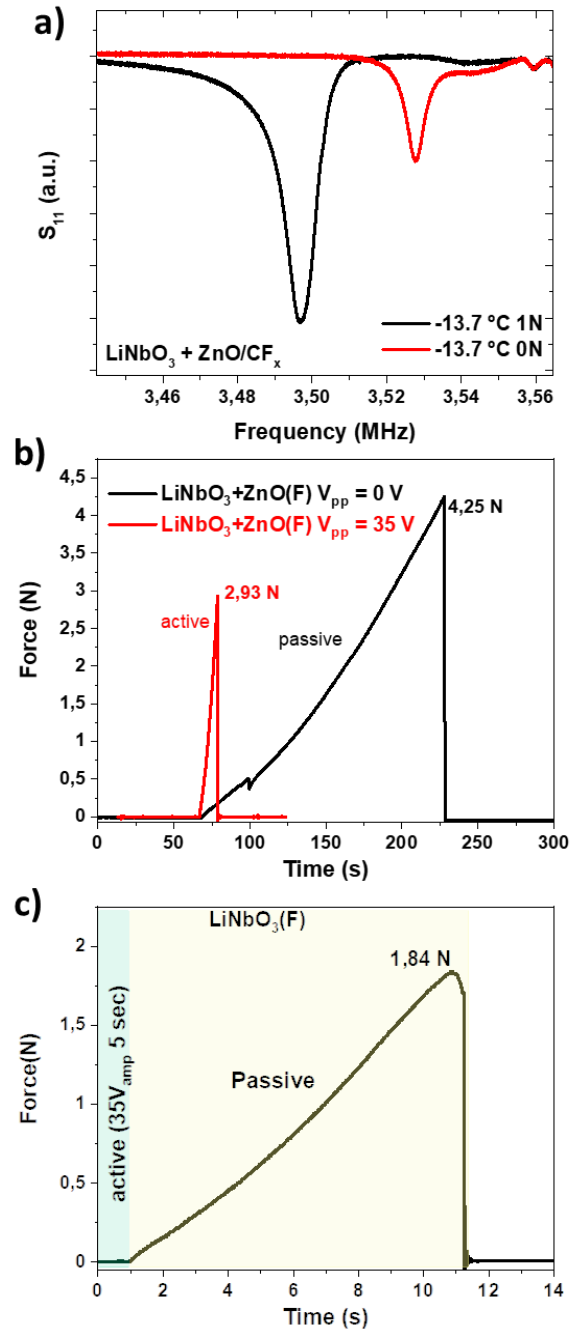

**Figure 5. Ice adhesion behavior under acoustic wave activation.** a) Return loss spectra for a sample of $LiNbO_3$+$ZnO/CF_x$ with the ice probe stuck on its surface without applying a force and while applying a pulling force of 1 N at the same temperature. b) Force-time curves recorded with the dynamometer for sample $LiNbO_3$/ZnO(F) without activation (reference experiment) and while applying a constant voltage of 35 V (i.e. under AW activation). c) Idem for sample $LiNbO_3$(F) combining AW activation and pulling sequentially: first the activation of the piezo plate at the resonance frequency (35V) for 5 seconds, then the pulling was carried out with increasing force until ice detachment.

As a summary of the experiments, **Table 1** gathers the adhesion strength values determined from the Force-time curves for a series of experiments carried out with $LiNbO_3$, $LiNbO_3$(F), $LiNbO_3$-ZnO, and $LiNbO_3$-ZnO(F) under AW actuation and pulling off. Values for reference experiments without AW activation are included for comparison. A first look at the reference samples (i.e. "passive" experiments) indicates that, in good agreement with the literature,[24,46,65,68] the hydrophobic character and the roughness are critical characteristics determining the adhesion strength. Thus, the adhesion strength is smaller for the hydrophobic $LiNbO_3$(F) sample than for the bare $LiNbO_3$ substrate, while the rougher $LiNbO_3$+ZnO(F) and $LiNbO_3$+ZnO samples present the highest adhesion strengths of the whole series. Much lower values of adhesion strengths were obtained while activating the substrates with AWs. Thus, most values reported in Table 1 are smaller than 100 kPa, a value usually taken as the threshold for low-adhesion substrates.[67] Moreover, data in this table show that the higher the applied voltage, the lower the adhesion strength. In some cases, ca. 20 kPa values were obtained that are characteristic of extremely low adherent surfaces.[67] In good agreement with the essays carried out on non-activated samples, for the actuated systems, hydrophobicity has a positive influence in reducing the adhesion strength (perfluorinated surfaces present lower adhesion values), while roughness (i.e. ZnO coated) increases it.

**Table 1.** Summary of the ice-adhesion results for the LiNbO3 and modified LiNbO3 chip plates. Passive essays were carried out on the same surfaces without applying the AW activation.

| Substrate | Voltage (AW) | Adhesion strength |
|---|---|---|
| $LiNbO_3$ (passive) | ---- | 134 kPa |
| $LiNbO_3$ | 20 V | 82 kPa |
| $LiNbO_3$(F) (passive) | ---- | 77 kPa |
| $LiNbO_3$(F) | 20 V | 29 kPa |
| $LiNbO_3$(F) | 35 V (5 s) | 20 kPa |
| $LiNbO_3$+ZnO (passive) | --- | 160 kPa |
| $LiNbO_3$+ZnO | 40 V | 140 kPa |
| $LiNbO_3$+ZnO(F) (passive) | --- | 150 kPa |
| $LiNbO_3$+ZnO(F) | 35 V | 71 kPa |

In summary, the set of adhesion strengths reported in Table 1 confirms that the AW activation of piezoelectric substrates decreases the ice adhesion, a feature contributing to an easier removal detachment of ice from surfaces.

### 3.4. De-icing and active anti-icing in an IWT

In this section, we check whether the AW activation of substrates contributes to decreasing the ice accretion capacity, as well to ascertain its efficiency to induce the *de-icing* of already accreted ice. Both sets of experiments were carried out under conditions close to those in real scenarios. For this purpose, we placed the plate holder in an IWT operated under the conditions described in the experimental section (see also Figure S5) and three air velocities: 25, 50, and 70 m/s. Previous to de-icing and ice accretion experiments, the response of the piezoelectric plate was monitored as a function of wind tunnel velocity. **Figure 6a)** shows the return loss spectra obtained for a bare $LiNbO_3$ plate placed in the wind tunnel exposed to dry air flows at -6ºC and the three velocities selected for the experiments. It is apparent that the resonance peak broadens and becomes displaced a few tenths of kHz when exposing the plate to increasingly higher air velocities. Although a detailed analysis of the factors contributing to this systematic variation is outside the scope of the present work (e.g. pressure drops due to Bernoulli effects or local temperature fluctuations might be involved in these variations), these findings further support the possibilities and high sensitivity of $LiNbO_3$ plates for environmental monitoring.

Similar to freezing experiments reported in Figure 3, changes during the icing process were also monitored in the IWT following the changes in the return loss spectra. An example of the icing monitoring capacity of the device is reported in the Supporting Information, **Figure S10**, showing the return loss spectra recorded in static conditions of the IWT (i.e. in the absence of wind) at -6º and -6.5ºC for a $LiNbO_3$ plate with a supercooled water droplet on its surface and the ice aggregate resulting from the droplet freezing, respectively. Return loss spectrum after freezing in the IWT depicted a considerable decrease in the intensity, a widening, and a clear shift in the frequency of the resonance maximum. From the perspective of the AW activation of the devices in the IWT under operating conditions (i.e. a mixture of supercooled water droplets in the airflow), the observed shift of the resonance frequency with the wind speed (Figure 6a) and the water-ice transformation (**Figure S7**) makes it compulsory to tune the resonance peak before and during the icing experiments. This strategy, explained in the experimental section, was systematically applied during the realization of the experiments reported next.

*3.4.1 Active anti-icing in IWT*

The first set of results deals with the *active anti-icing* capacity (ice accretion prevention) of the AW-activated piezoelectric plates. **Figure 6b)** shows an image taken in the IWT after an ice accretion test on the bare piezoelectric plate activated substrate (applied voltage 30 V) and on a reference LiNbO$_3$ sample. The image taken after 9 min shows that a big amount of ice becomes accreted on the reference plate, but no ice appears on the AW-activated device. A detailed evolution of the accumulation of ice on the reference substrate and the occasional formation of small droplets of water on the activated substrates during this experiment can be seen in the snapshots provided in **Figure 7** and **Video S4**. It is noteworthy that during this experiment the temperature was continuously monitored by an optical fiber sensor placed beside the piezoelectric plate (Figure 6b) and by the thermographic camera and that their maximum readings were during AW activation of the plate. The found effectiveness of the AW activation to prevent the formation of ice must be the consequence of various factors such as the low adhesion of ice on the AW activated substrates (see Table 1) and their capacity to induce the melting of small droplets of water (Figure 4). The remarkable final effect is that no ice became accreted on the activated substrate while the full area on the reference substrate sustained by the slit entrance of the shield (see Figure S5) was covered by glace ice.

According to **Video S4**, the monitoring of temperature with the optical fiber showed an increase maximum of -3 ºC, while the thermocamera variations were higher, in agreement with a larger area of inspection (please note that in Video S4 the higher temperatures detected by the photocamer corresponds to the areas beneath the metal electrodes which appear glowing during the tunning of the resonance frequency).

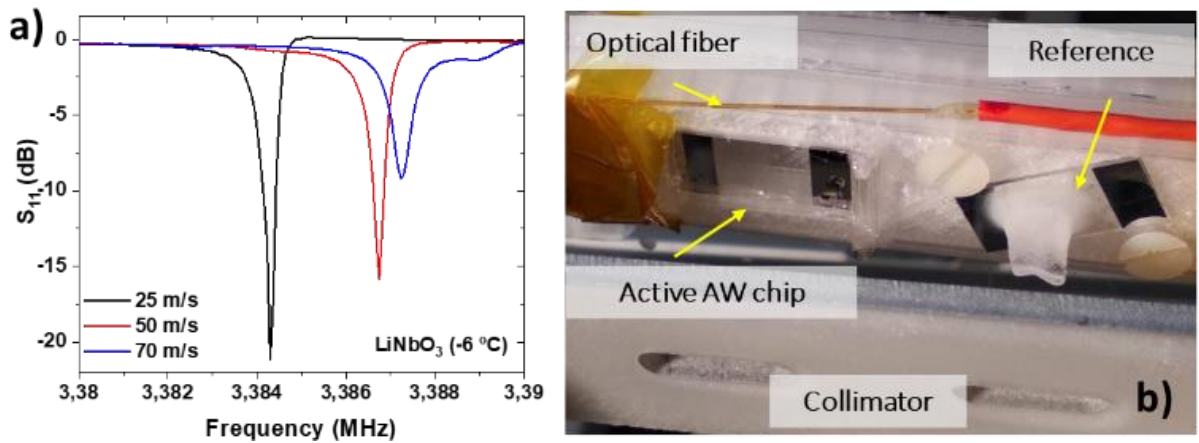

**Figure 6. Results of active anti-icing test in an IWT**. a) Return loss spectra recorded for a LiNbO$_3$ plate at – 6ºC in dry conditions under the action of different wind speeds as indicated. b) Photograph of the sample holder emplaced in the IWT behind the collimator after an active

anti-icing experiment where a high amount of glace ice became accreted on the reference sample. The AW activated chip remained ice-free. The main elements appear labeled in the photograph.

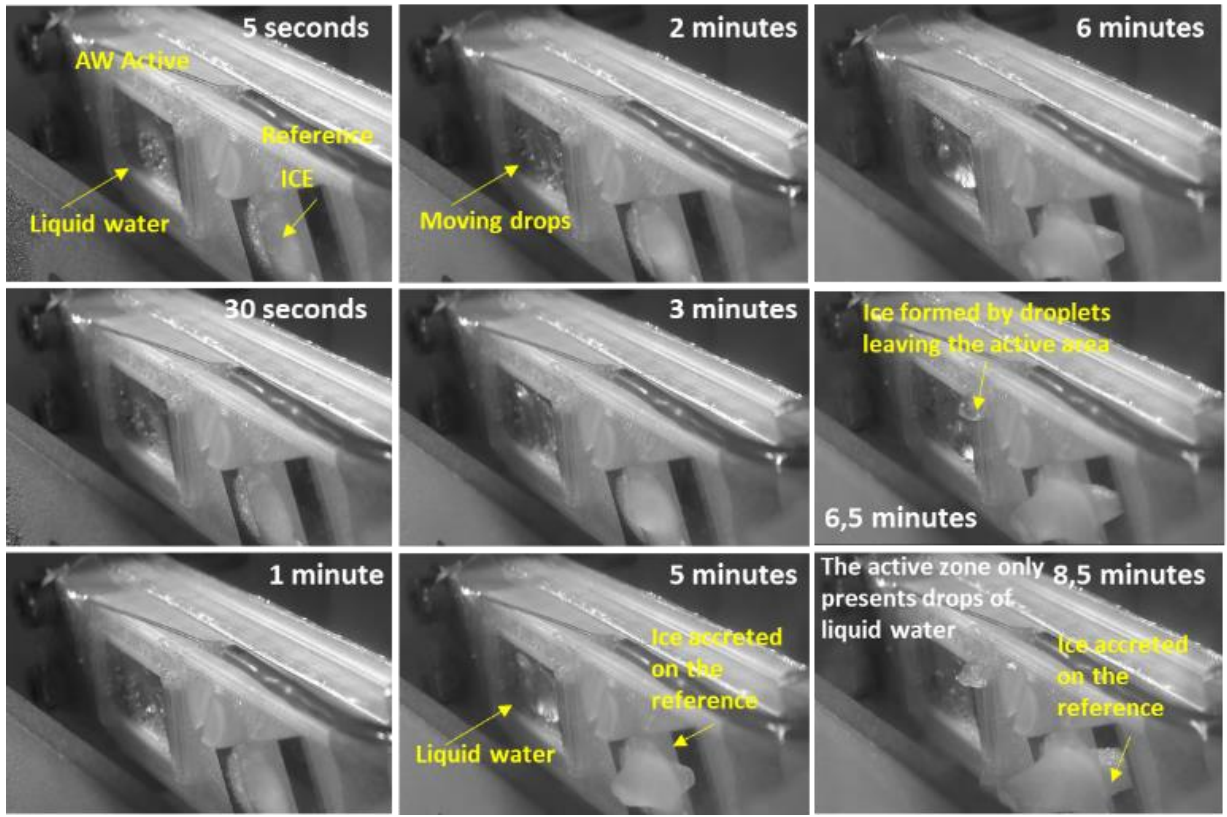

**Figure 7. Characteristic active anti-icing experiment carried out in the IWT under realistic conditions**. Successive snapshots from an ice accretion experiment revealing the active anti-icing capacity of the AW to prevent the formation of ice aggregates onto its surface for a bare LiNbO$_3$ plate. The left and right sides of the pictures correspond to the AW activated and reference sample correspondently. Maximum amplitude 30 V. IWT conditions: Glaze ice, v = 25 m/s, SAT = - 8 ºC; LWC = 0.5 g/m$^3$, MVD = 20 µm

*3.4.2 De-icing in IWT*

In the absence of AW activation ice became equally accreted on the reference sample and the chip device. *De-icing* experiments could then be carried out under the working conditions of the IWT. The series of snapshots in **Figure 8** and **Video S5** show that ice could be effectively melted and removed by the wind flow from the surface of the LiNbO$_3$ chip plates when activating the piezoelectric device with AWs. This experiment was carried out applying 30 V input voltage, after ice formation for the IWT operated at a wind velocity of v = 70 m/s and a temperature of -8 ºC. Remarkably, ice remained unaltered on the reference sample, but on the AW activated sample ice first changed opacity just after 7.5 s of applying the AWs and became molten after 8 secs. The resulting water droplets were dragged out the surface of the AW activated piezoelectric plate, which appeared completely dry just after 15 s. For the experiments in Figure 8, the optical fiber sensor measured a maximum temperature of -5 ºC just at the moment of the melting, the synchronized Video S5 captures such an instant showing a sudden increase of a few degrees in both the optical fiber and the thermocamera.

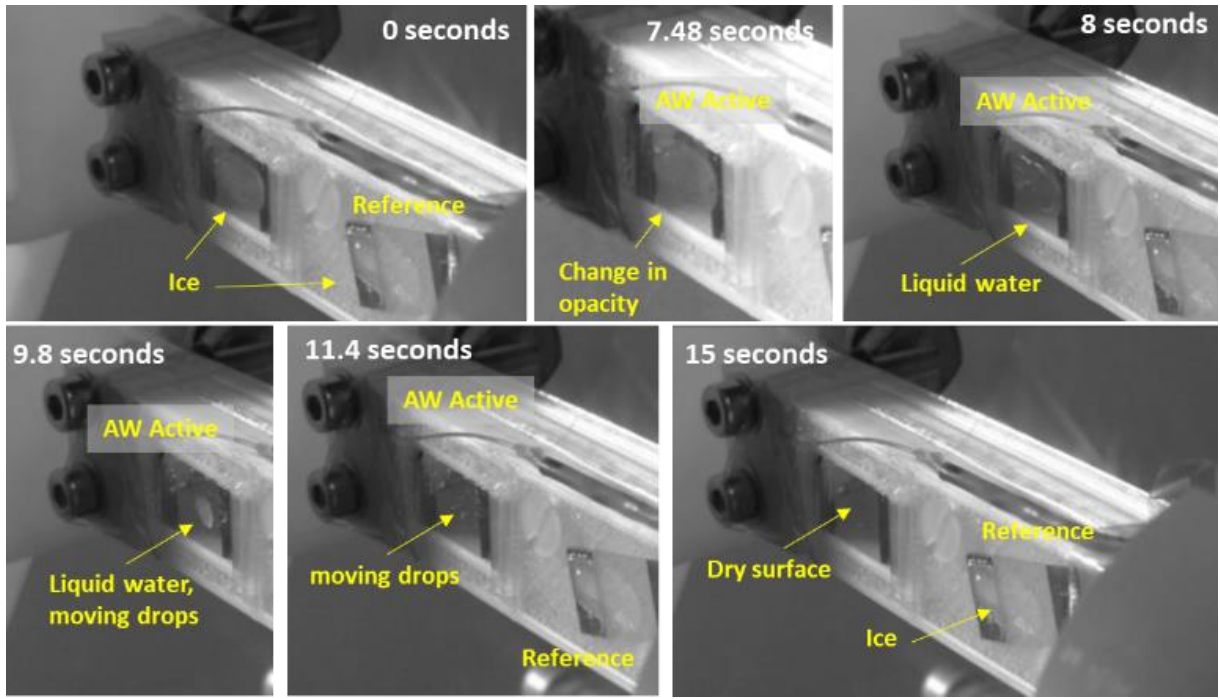

**Figure 8. Characteristic de-icing experiment carried out in the IWT under close to real conditions**. Successive snapshots from an ice accretion experiment showing the fast de-icing by AWs of the ice covering the LiNbO3 chip plate. Maximum amplitude 30 V. IWT conditions: Glaze ice, v = 70 m/s, SAT = - 8 ºC; LWC = 0.5 g/m3, MVD = 20 um)

Both the *de-icing* and *active anti-icing* experiments described above were carried out with bare LiNbO$_3$ plates, as well as for surface modified substrates LiNbO$_3$+ZnO(F) and LiNbO$_3$+ZnO/CF$_x$. **Table 2** gathers the applied voltages required to achieve equivalent *de-icing* and *active anti-icing* efficiencies for various samples and wind velocities from 0 to 70 m/s. In all cases, the optical fiber sensor renders values similar to those provided during the experiments reported in Figures 6-8. The series of voltage values in Table 2 can be used as an indicative parameter of the efficiency of the respective process under investigation, although conclusions should be taken on a semiquantitative basis because some parameters cannot be precisely determined under the IWT working conditions (e.g. the actual volume of accreted ice for *de-icing* experiments or amount of ice accreted on the reference sample for the active anti-icing tests). With these restrictions, the series of voltage values gathered in Table 2 indicate that the surface-modified substrates, characterized by a hydrophobic behavior, required less power than the bare substrates for equivalent *de-icing* or *active anti-icing* efficiencies. It can be also realized that, in most cases, the required power increased for higher wind velocities. This behavior is clear for *de-icing*, while for the active anti-icing experiments no neat tendency could be deduced, probably because the applied power likely surpassed in some experiments the minimum threshold power required to prevent icing.

**Table 2.** Summary of requiered voltages for selected de-icing and active anti-icing experiments in the IWT facility

| De-icing experiments: AW activation of the LiNbO$_3$ chip plate with ice formed on its surface | | |
|---|---|---|
| **Sample** | **V$_{amp}$ at Wind speed at 0 m/s** | **V$_{amp}$ at Wind speed at 70 m/s** |
| LiNbO$_3$ | 25 V | 30 V |
| LiNbO$_3$ + ZnO(F) | 10 V | 15 V |
| LiNbO$_3$ + ZnO/CF$_x$ | < 10 V | 15 V |
| **Active anti-icing experiments: AWs activated substrates during the arrival of supercooled droplets** | | |
| **Sample** | **V$_{amp}$ at Wind speed at 25 m/s** | **V$_{amp}$ at Wind speed at 70 m/s** |
| Transparent LiNbO$_3$ | 30 V | 25 V |
| Transparent LiNbO$_3$ + ZnO(F) | 17 V | 20 V |
| Transparent LiNbO$_3$ + ZnO/CF$_x$ | 17 V | 15 V |

These pieces of evidence taken under close to real environmental conditions confirm the positive effect of hydrophobicity on the AW *de-icing* and *active anti-icing*. For *de-icing*, high wind velocities decrease the efficiency, a feature that still requires further analysis considering factors such as actual pressure at the position of the plates and the actual value of water vapor pressure in contact with the accreted ice (water vapor would be dragged away more efficiently at high wind velocities), a variable that is known to contribute to the efficiency of *de-icing* under Joule well-controlled laboratory conditions.[69]

## 3. Conclusions

In this article, we have addressed a series of fundamental effects involved in the *de-icing* and related ice removal or *active anti-icing* processes induced by AW excitation of piezoelectric plates. To get information on these issues, experiments have been performed with well-defined piezoelectric substrates consisting of $LiNbO_3$ single crystal plates where standing acoustic plate waves with a majority shear character are generated by rf activation. The wetting properties of this ideal system could be readily modified by surface functionalization without altering the AW resonance conditions. In particular, we have demonstrated that the modification of the plate with a ZnO(F) thin film or $ZnO/CF_x$ multilayer renders the substrate surface hydrophobic while leaving unaffected the AW excitation. To the best of our knowledge, this is the first example demonstrating the *de-icing* capability of AWs in a multilayer anti-icing system. Besides, our results have provided a well-founded assessment of the influence of the wetting behavior and surface roughness of the plate surface on phenomena such as the adhesion and accretion of ice, or the ice melting phenomena. In general, surface hydrophobicity favors the anti-icing character of the piezoelectric chip devices, reducing the AW power required to induce *de-icing* or prevent icing (*active anti-icing*). Regarding the ice adhesion, a higher roughness of the functionalized devices appears to be detrimental for an effective detachment and contributes to increasing the adhesion strength of ice onto the modified piezoelectric substrates. Comparison between standard (transparent) and non-pyroelectric (black) $LiNbO_3$ has served to discard that, to a first approximation, pyroelectric activity plays a significant role in the AW *de-icing* processes, thus supporting previous hypotheses relying on pure mechano-acoustic mechanisms for the melting of ice aggregates by AWs.

Particular attention has been paid in this work to the freezing/melting behavior of water droplets. The study of an idealized system consisting of the $LiNbO_3$ piezoelectric plate has revealed important clues regarding the melting of ice aggregates under the effect of AWs activation. Specifically, our experiments have shown the ice-particle interface melting constitutes the first stage in the overall *de-icing* process for the wave mode studied here. For intermediate states, with both water and ice coexisting in partially melted aggregates, water seems to effectively wet the remaining ice core, providing a liquid diffusion layer over which the ice core can slide and be naturally removed by the action of natural forces such as gravity or wind (this latter as demonstrated in the icing wind tunnel experiments). Whether the shear character of the AWs employed in this work is critical for this mechanism is a question to be disclosed in the future with experiments where the AWs have a predominant out-of-plane component.

A relevant characteristic of the piezoelectric plate devices in the experiments carried out in this work is the high sensitivity of the resonance frequency of the plate in the utilized device to environmental and working parameters. In principle, this dependence might be considered a drawback in relation to the less sensitive thin film piezoelectric surface acoustic wave (SAW) systems operated with interdigitated electrodes at much higher frequencies. However, this high sensitivity of the resonance of piezoelectric plates towards environmental conditions (temperature, pressure, formation of ice, effect of wind, etc.) can be turned into an advantage for environmental sensing, provided that an effective tuning of the resonance frequency can be applied during the icing experiments. In this regard, we have proposed that the preliminary results shown in this work about the variation of the resonance frequency with environmental variables may set the basis for the development of specific sensing devices with a simpler architecture than state-of-the-art SAW devices relying on thin films and IDTs.

**Supporting Information**

Supporting Information is available from the Wiley Online Library or from the author.

**Acknowledgements**

We thank projects PID2019-110430GB-C21, PID2019-109603RA-I00, and PID2020-112620GB-I00 funded by MCIN/AEI/10.13039/501100011033 and by "ERDF (FEDER) A way of making Europe", by the "European Union". The project leading to this article has received funding from the EU H2020 program under grant agreement 899352 (FETOPEN-01-2018-2019-2020 - SOUNDofICE).

Supporting Information

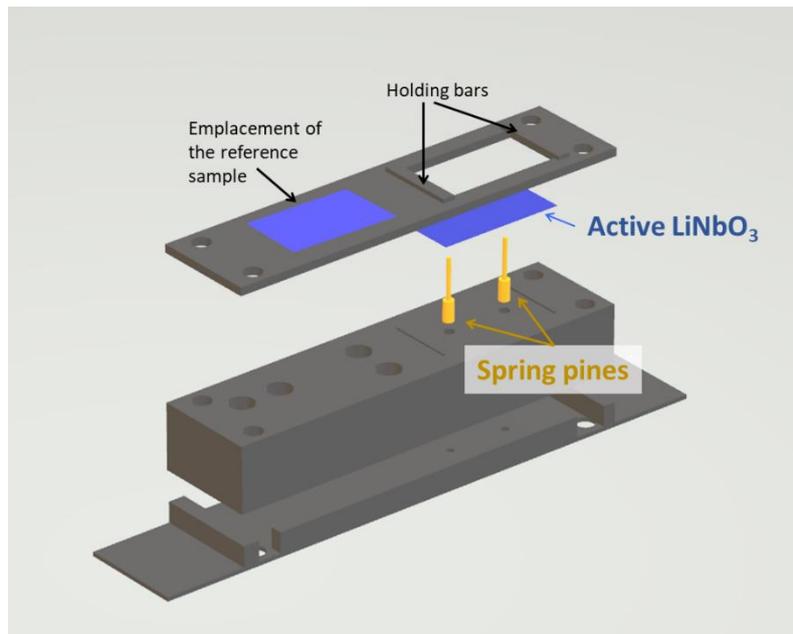

**Figure S1.** Detailed schematic of the holder design for the de-icing by acoustic waves employing piezoelectric plates.

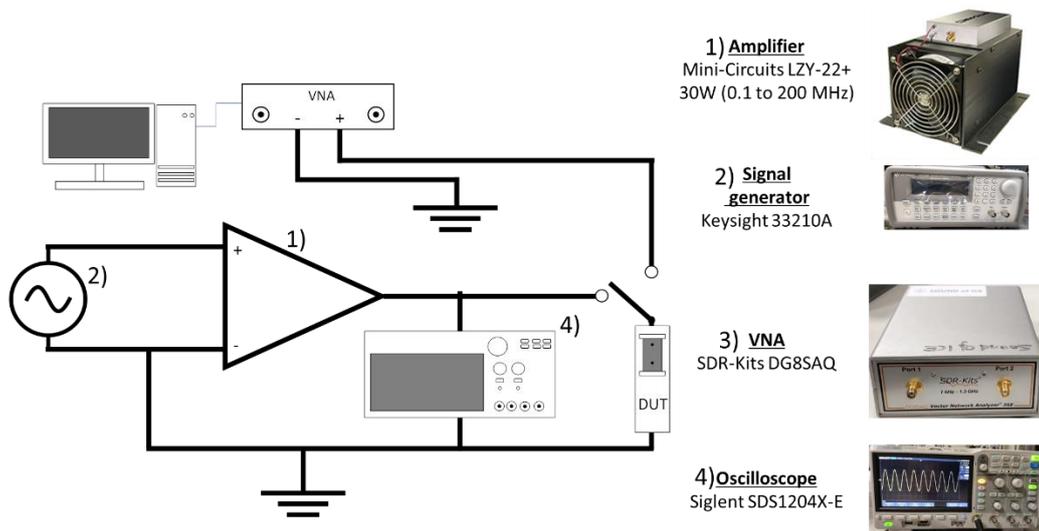

**Figure S2.** Schematics of the excitation and syntonization circuits for the actuation of the de-icing AW devices.

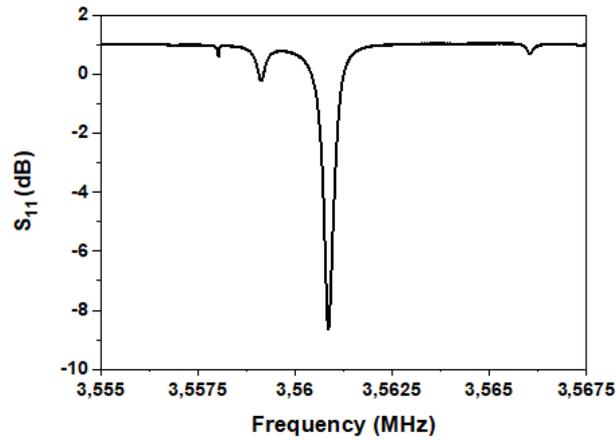

**Figure S3.** Experimental $S_{11}$ for a LiNbO$_3$ plate (2 x 1 cm$^2$) actuated by two extended Ta electrodes (1 x 0.5 cm$^2$)

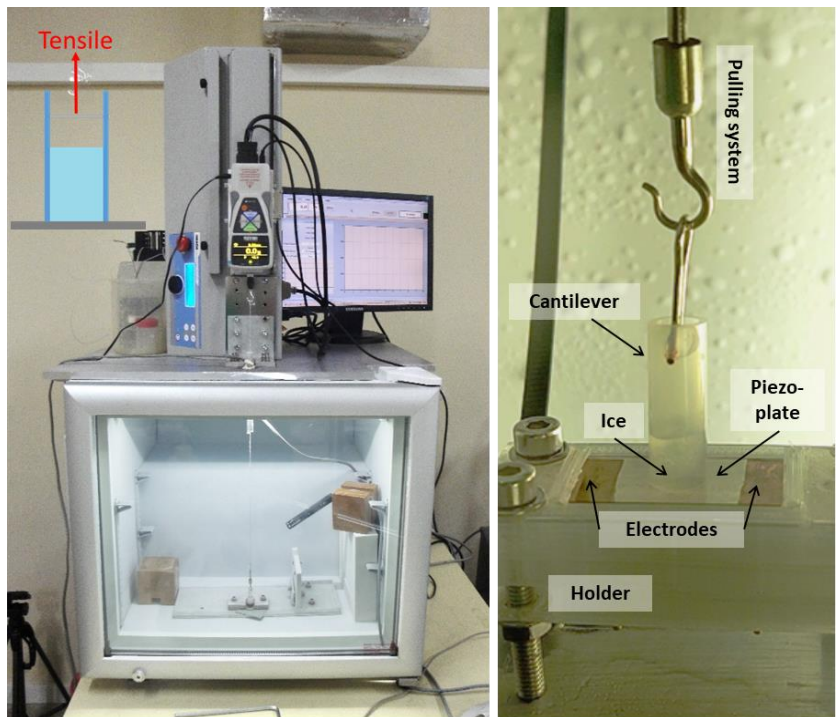

**Figure S4.** Photographs of the pull-off ice-adhesion characterization test (left) and details of the cantilever placed on top of the AW de-icing device (right)

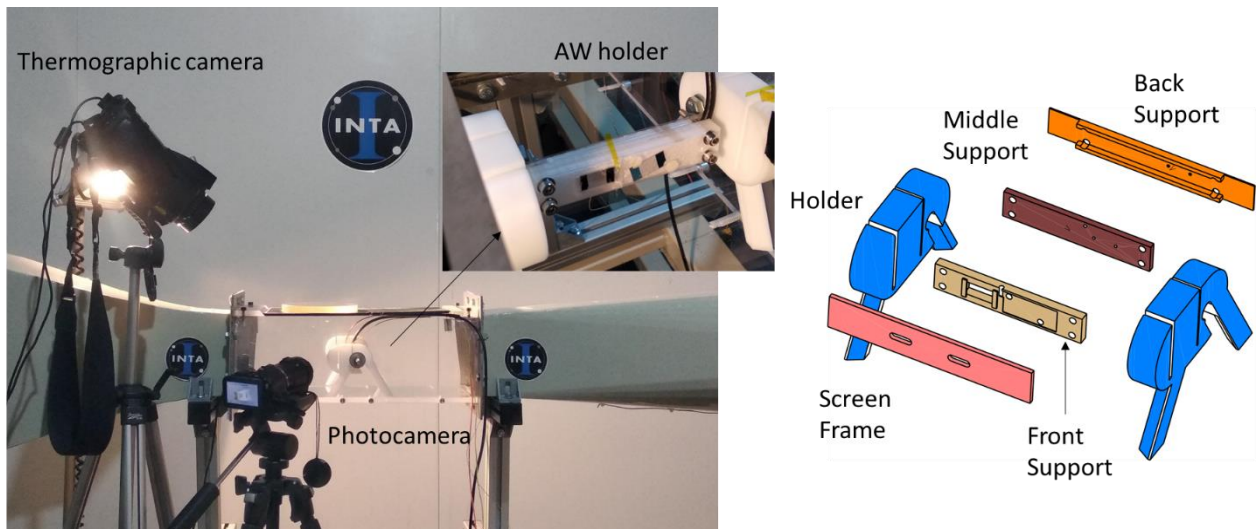

**Figure S5.** Photographs (left) and schematics (right) of the experimental setup employed in the IWT for the simultaneous characterization of AW actuated and reference samples.

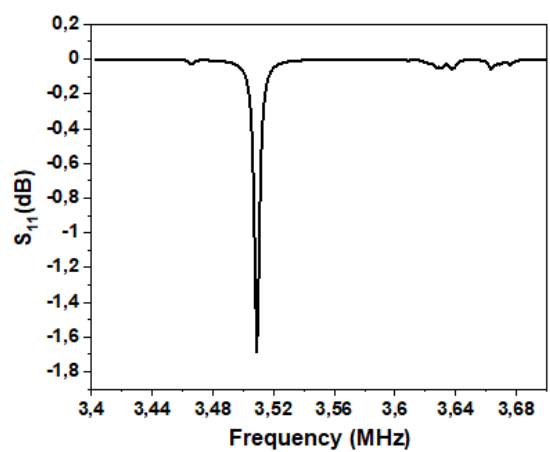

**Figure S6.** Simulated $S_{11}$ for a LiNbO$_3$ plate (3 x 1.5 cm$^2$) actuated by two extended electrodes (3 x 0.5 cm$^2$).

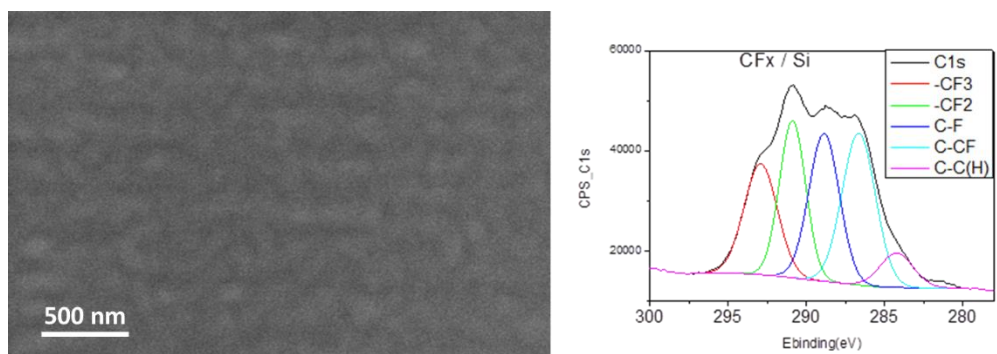

**Figure S7.** Left) SEM normal view micrograph of a CF$_x$ thin film deposited on ZnO; Right) XPS zone peak corresponding to C1s for a CF$_x$ on Si.

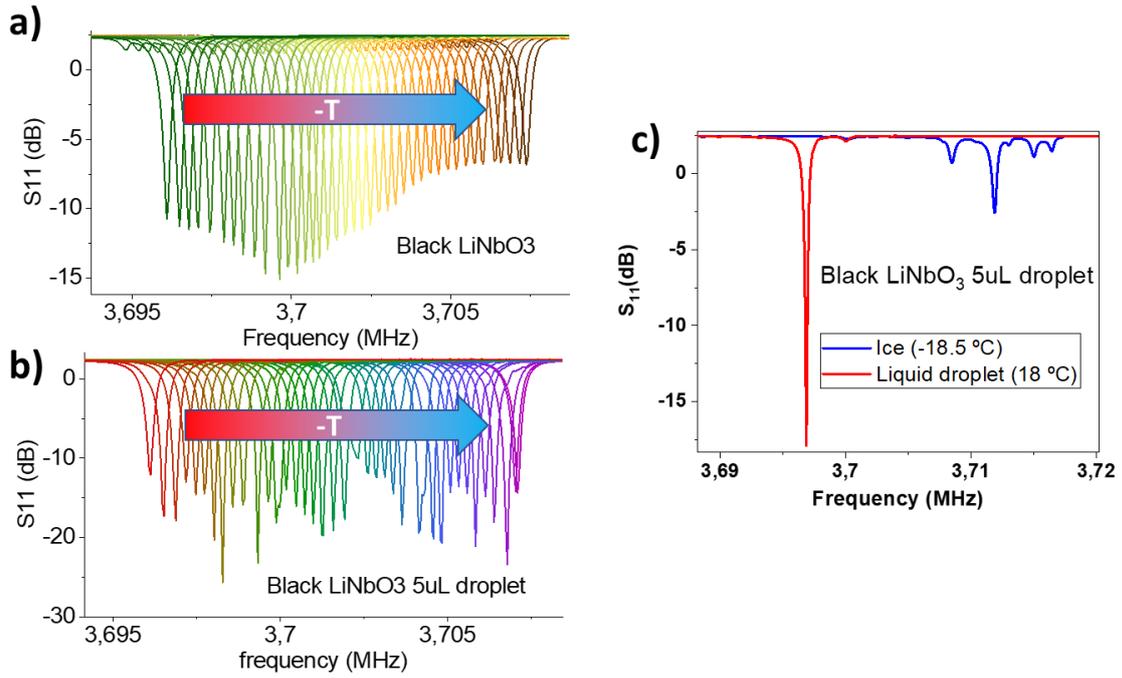

**Figure S8**. Variation of return loss spectra of a black LiNbO$_3$ plate with planar electrodes, as a function of temperature (a) and in the presence of water droplet (b). c) Comparison of the S$_{11}$ under liquid and ice water conditions.

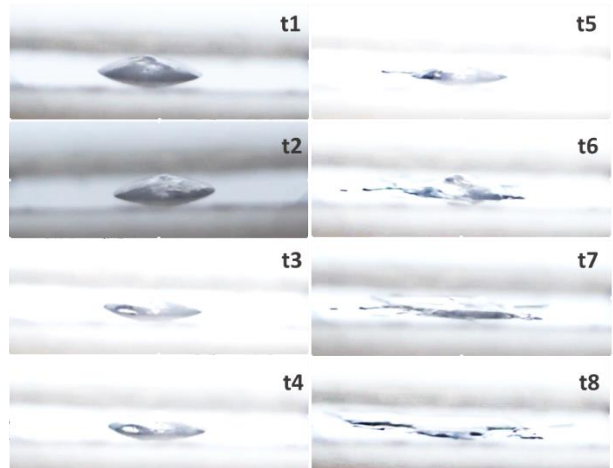

**Figure S9.** Snapshots for the melting of ice aggregates onto black LiNbO$_3$ substrates.

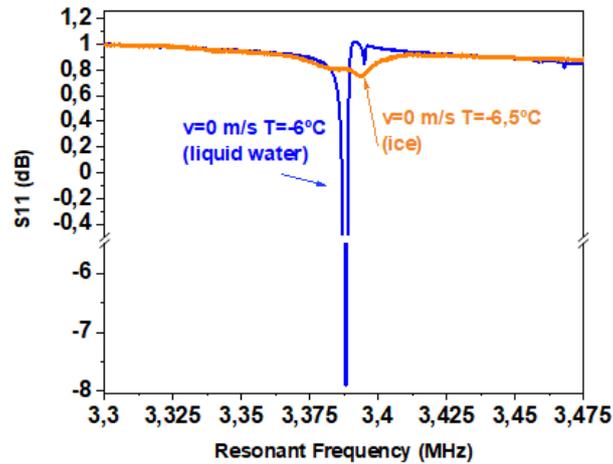

**Figure S10.** Return loss spectra for a liquid (blue) and frozen droplet (orange) on the surface of a bare LiNbO$_3$ recorded at the IWT.

**Video S1.** Animation representing the acoustic wave actuation on LiNbO3 by extended electrodes obtained from FEM.
**Video S2.** Recording obtained in the OCA WCA system of the comparison of the de-icing by acoustic waves (left) and Joule effect (right) on a flat LiNbO$_3$ surface.
**Video S3.** Recording obtained in the OCA WCA system of de-icing by acoustic waves on a LiNbO$_3$ surface tilted 90º.
**Video S4.** Results of de-icing by AWs on LiNbO$_3$ corresponding to Figure 7: Photocamera (left), thermocamera (middle), and optical fiber (right) emplaced in the IWT.
**Video S5.** Results of active anti-icing by AWs on LiNbO$_3$ corresponding to Figure 8: Photocamera (left), thermocamera (middle), and optical fiber (right) emplaced in the IWT.